\documentclass[12pt]{article}
\usepackage[utf8]{inputenc}
\usepackage{graphicx, rotating}
\usepackage{xcolor}

\usepackage[font=small]{caption}
\usepackage{caption}
\newenvironment{myminipage}[1]{\minipage{#1} 
\captionsetup{width=\textwidth, name=Fig., labelfont={it,bf}}
 }{\endminipage} 

\newcommand{\kp}{\mbox{\boldmath $k$}_{\perp }}

\newcommand{\pvet}{\mbox{\boldmath $p$}}
\newcommand{\kvet}{\mbox{\boldmath $k$}}

\newcommand{\be}{\begin{equation}}
\newcommand{\ee}{\end{equation}}
\newcommand{\ben}{\begin{displaymath}}
\newcommand{\een}{\end{displaymath}}
\newcommand{\sixe}{\mbox{$e^+ e^-\to 3(e^+e^-)$}}
\newcommand{\foure}{\mbox{$e^+ e^-\to 2(e^+e^-)$}}
                       
\title{Searching for dark sectors in multi lepton final state in $e^+e^-$ collisions}

\author{Paolo Ciafaloni$^1$, Gabriele Martelli$^2$, Mauro Raggi$^3$}
\date{%
    $^1$INFN - Sezione di Lecce and Universit\`a del Salento,\\ %
    $^2$Dipartimento di Fisica e Geologia, Universit\`a di Perugia,\\%
    $^3$Dipartimento di Fisica, Sapienza Universit\`a di Roma,\\
}

\begin{document}

\maketitle

\begin{abstract}
Electron positron collisions are a very promising environment to search for 
new physics, and in particular for dark sector related observables.
The most challenging experimental problem in detecting dark sector candidates
is the very high associated Standard Model background.
For this reason it is important to identify observables that are, at the same 
time, minimally suppressed in the dark sector and highly suppressed in the 
Standard Model. One example is the $\sixe$ process  
that can be mediated either by the production and subsequent decay of dark 
Higgs ($h'$), $e^+e^-\to A'h'\to 6e$ \cite{Batell:2009yf} or produced by the 
Standards Model process $\sixe$. In the following 
letter we study the relative contribution to observed $\sixe$ total cross section,
coming from the $h'$ mediated and from the Standard Model
processes in the contest of fixed target and low energy collider experiments, with particular attention to the PADME experiment 
at the INFN Laboratori Nazionali di Frascati.
\end{abstract}

\section{Introduction}

In recent years, the idea that dark matter can be considerably lighter than 
the weak scale and secluded by portal interactions, has attracted growing 
interest also supported by the very stringent limits on dark matter at 
GeV-TeV mass scales obtained by direct detection and LHC experiments. 

Among the possible different realisations of dark sectors involving portal 
interaction and new mediators, the kinetic mixing with U(1) hypercharge
is of considerable interest from a phenomenological point of view
as the channel where hidden sector may be probed with maximal sensitivity.

The authors of \cite{Batell:2009yf} have investigated the signatures of a
minimal secluded U(1)$_D$ extension of the SM at medium energy colliders while in \cite{Essig:2009nc} the dark group is non-abelian. 
The proposed models assume the existence, in addition to the U(1)$_D$ ``dark'' photon $A'$, of an elementary dark Higgs boson 
($h'$), which spontaneously
breaks the U(1)$_D$ symmetry, and of dark matter particles with WIMP scale 
masses. 

The amplitude of dominant production mechanism for the $h'$, the so called 
Higgs'-strahlung $e^+e^-\to A'h'$ 
\cite{Batell:2009yf}, is minimally suppressed in the dark sector by just a 
single power of the kinetic mixing parameter $\epsilon$. 

Under the assumption that both $m_{h'}$ and $m_{A'}$ are below the di-muon threshold 210MeV, at the end of the decay chains of both 
$h'$ and $A'$ we will only find SM particles

The $A'$ decay will produce an $e^+e^-$ pair while the $h'$ decay will produce  
to $2(e^+e^-)$, leading to a minimally suppressed final state of 3($e^+e^-$). 
On the contrary the Standard Model process $e^+e^-\to$3($e^+e^-$) suffers a strong $\alpha^6$ suppression, 
which can compensate the $\epsilon^2$ suppression of the dark sector rate.

Experiments at $e^+e^-$ colliders with few GeV center of mass energy and  
very high luminosity, Babar \cite{babar} and Belle \cite{belle}, had investigated $e^+e^-\to3(\ell^+\ell^-)$
setting stringent limits on the existence of the $h'$ for energy scales in 
the $\sim$GeV range. No data are available for masses of the $h'<$1 GeV which can
be probed by low energy fixed target experiments.

For this reason it is worth investigating the scale of the contribution of dark sector particles decays to the cross section of 
the process $\sixe$, compared to Standard Model one at low energy colliders. 
Being the dark contribution in any case small, in this paper we will develop new techniques to precisely calculate 
the SM contribution.

The paper is organised as follows: in section 2 (3) we discuss the various approximations for 
the  $e^+ e^-\to 2(e^+e^-)$  ($e^+ e^-\to 3(e^+e^-)$ ) process. In section 4 we introduce the Dark Higgs Model which is our reference one in this paper, and we compare $e^+ e^-\to 3(e^+e^-)$ cross sections in this model and in the Standard Model. In section 5 we specialise the discussion to the PADME experiment; our conclusions are drawn in section 6. Finally, Appendix A is devoted to the Equivalent Photon Approximation (EPA) which is used throughout the paper.

In the following we refer to $e^+ e^-\to 2(e^+e^-)$ as the ``$4\ell$ process'' and to 
 $e^+ e^-\to 3(e^+e^-)$ as the ``$6\ell$ process''. 


\section{$e^+e^- \to e^+e^-e^+e^-$ in the equivalent photon approximation and beyond}

Due to the large number of Feynman diagrams involved, there is no exact analytic expression for the  QED\footnote{At the energies considered here and for the given processes, the QED sector of the Standard Model is the only relevant one.} tree level cross section for $\foure$, nor for
$\sixe$. Therefore, one has to make use to various kinds of approximations, that follow three main lines:
\begin{itemize}
\item
The Equivalent Photon Approximation (EPA) \cite{Budnev:1974de}, similar to the parton model approach in QCD, which is described in detail in Appendix \ref{appA}.
\item
Tree level numerical calculations; here we perform such calculations making use of the CalcHEP tool \cite{Belyaev:2012qa}.
\item 
Approximate analytic expressions \cite{Budnev:1974de}, which retain the leading term in an expansion in the small parameter $m^2/s$, $m$ being the electron mass and $\sqrt{s}$ the c.m. energy.
\end{itemize} 
In the latter case, the analytic approximation has the form:
\begin{equation}
\sigma=\frac{1}{m^2}\left(a_n\log^n l^n+a_{n-1} l^{n-1}+\cdots +a_0\right)+{\cal{O}}(\frac{1}{s})
\end{equation}
where $n=3$ ($n=4$) for the 4$\ell$ (6$\ell$) process. 
The EPA allows to calculate the first term in the series, with the highest power of $l$; we call this the 'Leading Log EPA' (LLEPA)and we discuss how to obtain this approximation in Appendix \ref{appA}.

In the literature there are different analytical formulae which allow to compute the 
total cross section for the SM process $\foure$. 
We will start considering the calculation using the Equivalent Photon Approximation (EPA), and then compare it to 
more accurate calculation which are possible in the 4$\ell$ case.
According to a classical paper dated back to 1975 \cite{Budnev:1974de} the total 
cross section $e^+e^-\to e^+e^-e^+e^-$ using equivalent photon approximation at leading log 
reads:
\begin{equation}
    \sigma_{e^+e^-\to e^+e^-e^+e^-}\approx\frac{28\alpha^4}{27\pi m_e^2}\left(\log\frac{s}{m_e^2}\right)^3
    \label{eqn:4lEPA}
\end{equation}
where $s$ is the center of mass energy squared that for a fixed target experiment is $s \sim2E_{beam}m_e$.
Equation (\ref{eqn:4lEPA}) only includes the leading logarithmic term of the cross section while more accurate calculation can be obtained analytically. The paper by Budnev et al. \cite{Budnev:1974de} 
also contains, in its appendix F1, a complete analytical calculation of the cross section in which all the coefficients for the lower order logarithmic terms are worked out. 

\begin{equation}\label{llepa4l}
    \sigma_{e^+e^-\to e^+e^-e^+e^-}=\frac{28\alpha^4}{27\pi m_e^2}\left( \ell^3-A\ell^2+B\ell+C\right)+O\left(\frac{1}{s}\right) ;\qquad \ell=\log\frac{s}{m_e^2}
    \label{eqn:4lBGMS_full}  
    \end{equation}
where $\ell$ is the logarithmic term in Eq. \ref{eqn:4lEPA}.  
For electron-positron scattering the coefficients A,B,C have been calculated, neglecting interference between produced and scattered electrons, to be\cite{Budnev:1974de}:
\begin{equation}
A = 178/28 \sim 6.36 \qquad
B \sim  -11 \qquad
C \sim  100  
\end{equation} 
Looking at equation \ref{eqn:4lBGMS_full}, it appears that in fixed target experiment, in which $s$ is typically small and the $\ell$ term ranges from 7-10, none of the logarithmic term is negligible due to the high values of coefficients
A, B and C. At $e^+e^-$ colliders like PEP-II, KEKB and SuperKEKB, where $s$ can reach the 10 GeV range and the log term the value of 20, the $\ell^2$ terms still account for $\sim$30\% of the total cross section.   

Performing a numerical analysis we observe that, in particular at low energy colliders, accidental cancellations are almost complete among the dominant $\ell$ terms. 
At the energy scale of the PADME experiment ($E_B\sim$500MeV) we obtain:
\begin{itemize}
\item $\frac{28\alpha^4}{27\pi m_e^2}\ell^3       \sim  6.0\times10^8$  pb
\item $\frac{28\alpha^4}{27\pi m_e^2}(-A\ell^2) \sim -5.0\times10^8$ pb
\item $\frac{28\alpha^4}{27\pi m_e^2}(B\ell)      \sim -1.16 \times10^8$ pb
\item $\frac{28\alpha^4}{27\pi m_e^2}(C)          \sim1.4\times10^8$ pb
\end{itemize}
In this energy regime the sum of the three logarithmic terms of Eq. \ref{eqn:4lBGMS_full} almost cancels out and only accounts for $\sim15\%$ of the total cross section, which is surprisingly dominated by the constant term.  
For this reason is crucial to check, in the absence of experimental results to compare with, if the coefficients
 A,B, and C of Eq. (\ref{eqn:4lBGMS_full}) have been correctly calculated.

\subsection{Calculation of $\sigma(e^+e^-\to e^+e^-e^+e^-)$ using CalcHEP}

In order to check the reliability of Eq. (\ref{eqn:4lBGMS_full}) we have performed a tree level numerical calculation of the cross section of the process $\foure$ based on the CalcHEP tool \cite{Belyaev:2012qa}.
To improve the precision of the calculation, the mass of the electron was introduced in the CalcHEP model and the value
of the electromagnetic constant $\alpha$ adapted to low energy scales. The contributions coming from diagrams mediated by the Z bosons, which have been verified 
using CalcHEP to be smaller than $1\times10^{-4}$, are neglected in the calculations. 

\begin{figure}[htpb] 
\begin{myminipage}{0.40\textwidth}
\includegraphics[height=5cm]{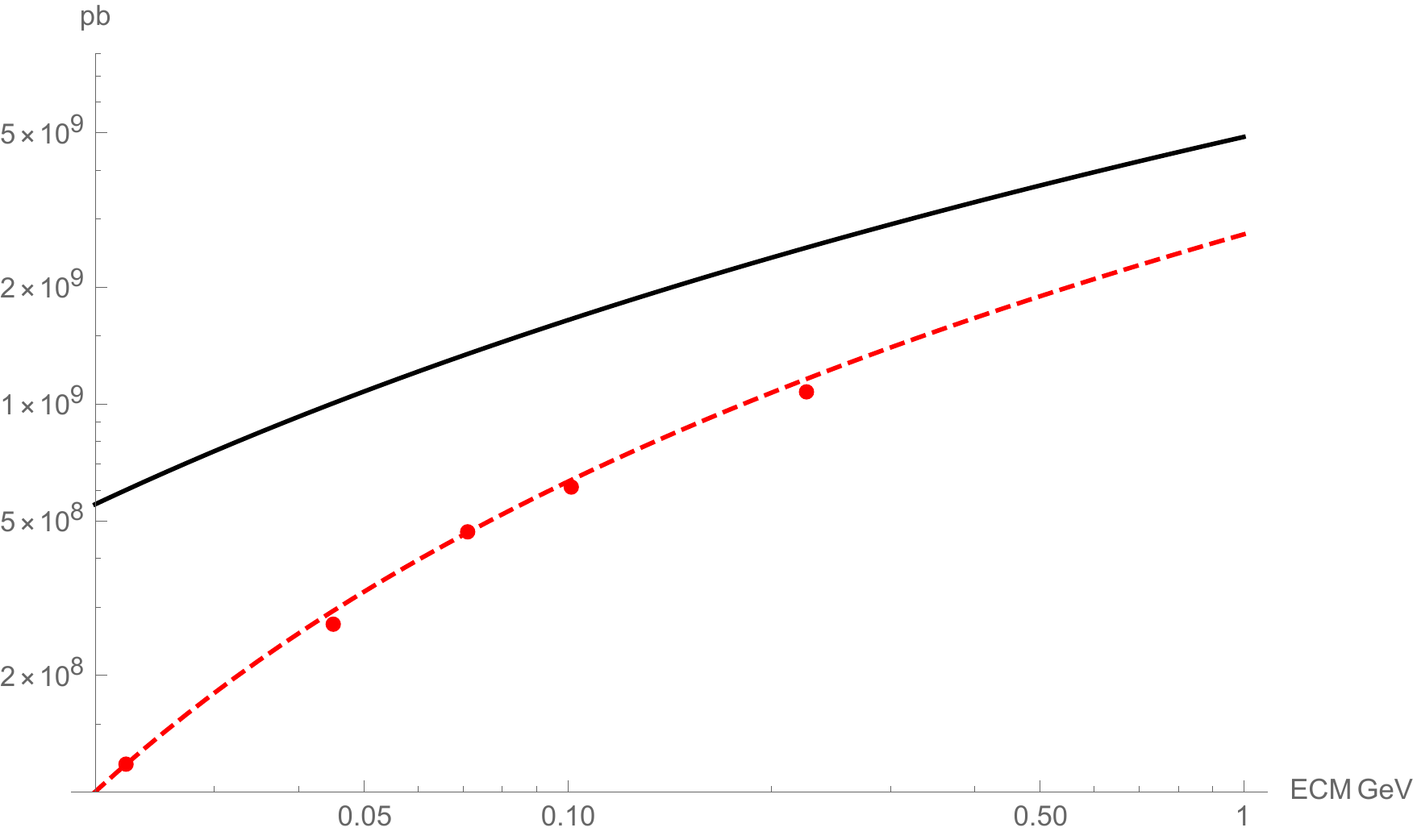}\\
\caption{{\em Comparison of $\sigma_{e^+e^-\to e^+e^-e^+e^-}$ at fixed target CM energies computed with Eq. (\ref{eqn:4lBGMS_full}), dashed red line, with CalcHEP, red dots, and with Leading Log EPA Eq. (\ref{eqn:4lEPA}), black line.}}
\label{fig:CompFT}
\end{myminipage} 
\hspace{1cm} 
\begin{myminipage}{0.40\textwidth}
\includegraphics[height=5cm]{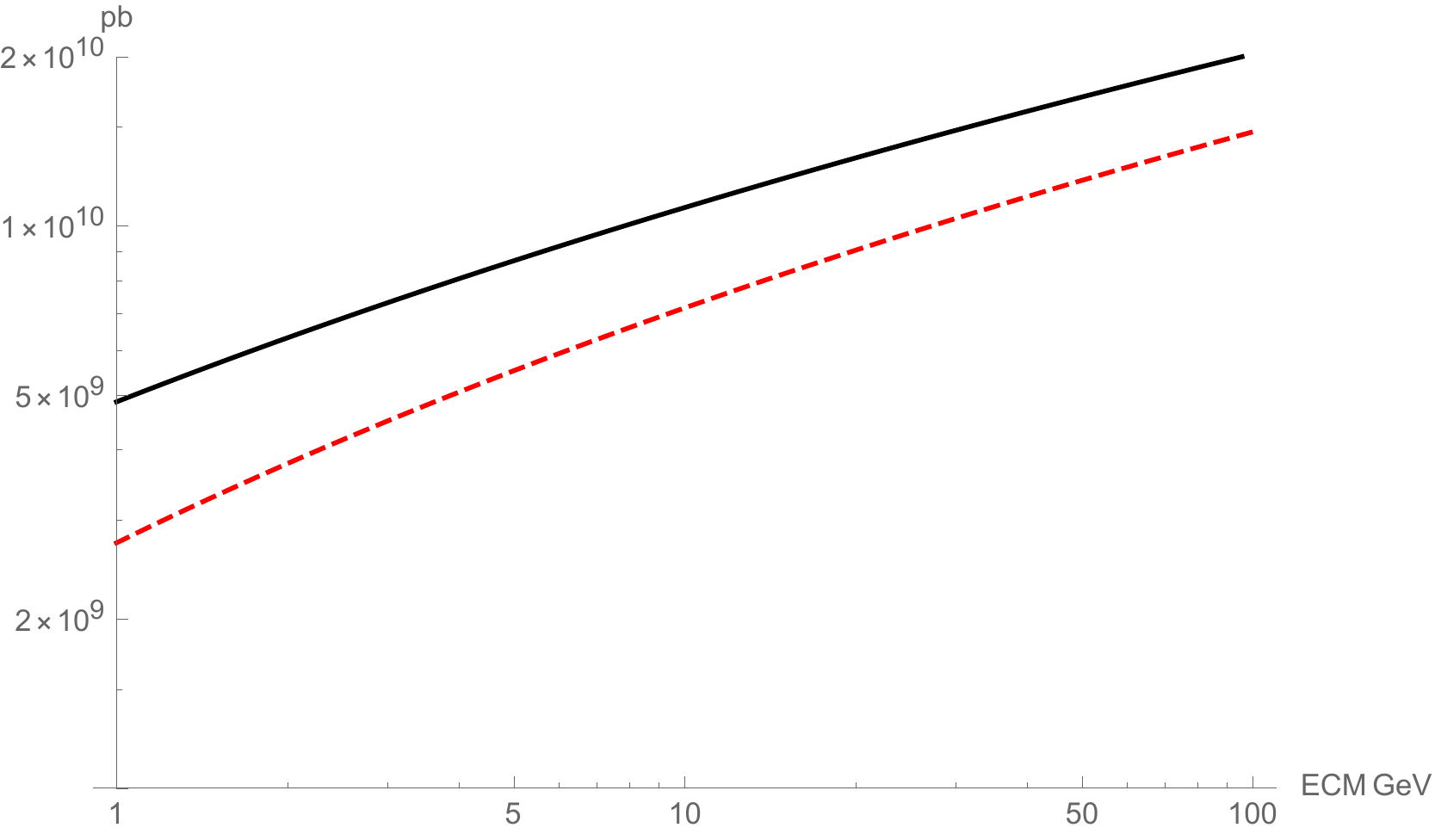}\\
\caption{{\em Comparison of $\sigma_{e^+e^-\to e^+e^-e^+e^-}$ at collider CM energies computed with eq (\ref{eqn:4lBGMS_full}), dashed red line, with CalcHEP, red dots, and with leading log EPA eq (\ref{eqn:4lEPA}), black line.}}
\label{fig:CompColl}
\end{myminipage} 
\end{figure}
 
In Fig. \ref{fig:CompFT} the comparison is shown for different values of the center of mass energy corresponding to fixed target experiments at different possibile extracted positron beam lines.
The first point corresponds to the DA$\Phi$NE Linac beam energy 0.5 GeV, the third point corresponds to the energy of an extracted positrons beam from the Cornell Electron Storage Ring (CESR 5 GeV) , 
the fourth point, beam energy $\sim$10 GeV, corresponds to the proposed JLab positron facilities at the CEBAF accelerator \cite{Accardi:2020swt}, while the last one at 50 GeV a positron beam produced by the North area facilities at CERN.
The plot show that a very good agreement, to the level of few \%, is achieved at any energy in between the fully analytical cross section and the CalcHEP based tree level calculation, while the leading log EPA approximation 
is inaccurate by factors, from 5 to 2. This huge discrepancy below the GeV scale originates from an accidental cancellation between the $\ell^3$ and the $\ell^2$ terms in Eq. (\ref{llepa4l}); moreover, the constant term is $\sim 100$. 
Of course, at higher energies the $\ell^3$ term starts to dominate and the Leading Log EPA becomes a better approximation of the analytic expression.

In Fig. \ref{fig:CompColl} the comparison is shown for different $s$ values relevant for collider experiments from KLOE at DA$\Phi$NE, 1GeV, to the experiments at the CERN LEP collider, 200 GeV. 
In this regime the overestimate of the cross section value due to the Leading Log EPA approximation, $\sim$ 2, appears to be more stable over a large range of energy, as the cross section also is.  To conclude, the leading log EPA provides a poor approximation of the analytic expression for the cross section of the $4\ell$ process due to accidental cancellations.

\subsection{Calculation of $\sigma_{e^+e^-\to e^+e^-e^+e^-}$ using numerical EPA approximation}

As shown in the previous paragraph the precision of the leading log EPA approximation in predicting the value of the $\sigma_{e^+e^-\to e^+e^-e^+e^-}$ is poor, in particular at energies below the GeV.
For this reason we developed an alternative approach which allows to maintain better precision, without performing a full tree level calculation as we did with CalcHEP.
This new technique will be very useful to estimate the $\sigma_{e^+e^-\to e^+e^-e^+e^-e^+e^-}$ for which a full tree level calculation cannot be achieved.

\begin{figure}
\begin{center}
\includegraphics[width=8 cm]{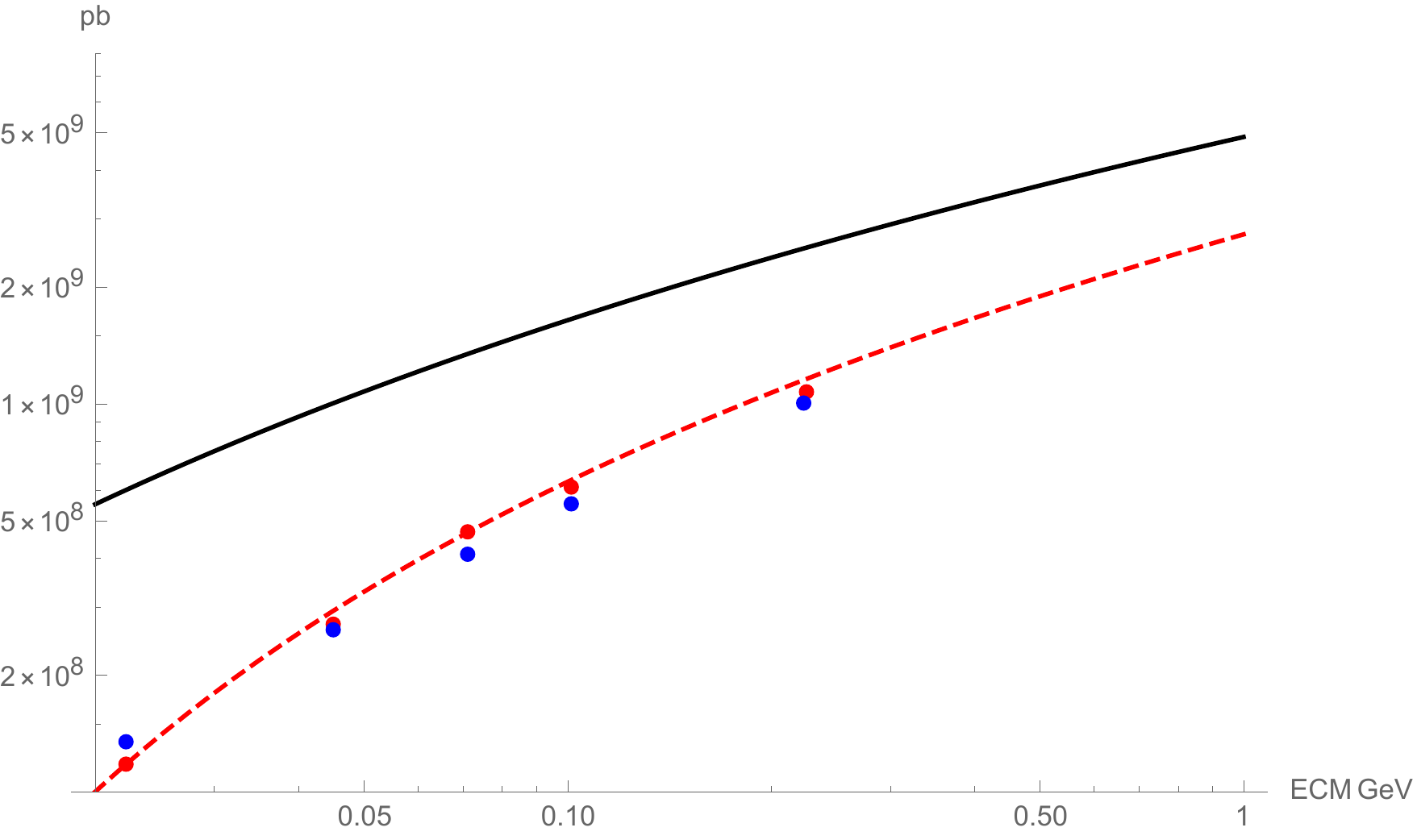}\\
\caption{{\em Comparison of $\foure$ at fixed target CM energies. Analytic calculations:  full EPA Eq.\ref{eqn:4lBGMS_full}, dashed red line, LLEPA Eq.\ref{eqn:4lEPA}, black line. Numerical calculations: Calchep, red dots, ``numerical EPA'', blue dots.}}
\label{fig:ImprovedEPA}
\end{center}
\end{figure}

Let us summarise the EPA with the following expression (\ref{lumi}):
\be\label{lumia}
\sigma(s)=\int_\epsilon^1d\tau{\cal L}(\tau)\sigma_p(\tau s)\qquad
{\cal L}(\tau)=\int_\tau^1\frac{dx}{x}f_{\gamma e}(x)f_{\gamma e}(\frac{\tau}{x})
\ee
In the leading log EPA we approximate $f_{\gamma e}({x})$ and $ {\cal L}(\tau)$ in order to obtain the leading log value for the cross section, as explained in the Appendix. 
However we can instead retain the full expressions for the splitting functions and the luminosity, thus improving the approximation. The full expression for the splitting function can be obtain by integrating inm $\kp^2$ the expression
given in eq. (5.18) of \cite{Budnev:1974de} from $|\kp^2|_{min}=x^2m^2$ to $|\kp^2|_{max}=m^2$ to produce:
\be\label{skkf}
f_{\gamma e}(x)=\frac{\alpha}{2\pi}\left(\frac{1+(1-x)^2}{x}\log\frac{1}{x^2}-(1-x)(1-x^2)\right)
\ee
while the expression for the cross section $\gamma \gamma\to e^+e^-$ can be found in (E.4) of  \cite{Budnev:1974de}:
\be\label{porcu}
\sigma_{\gamma\gamma}=\frac{4\pi\alpha^2}{s\tau}\left((1+4\frac{m^2}{s\tau}-8\frac{m^4}{s^2\tau^2}L-(\frac{1}{\tau}+4\frac{m^2}{s\tau^2})t\right)
\ee
with
\be\label{porcu2}
L=2\log\left[\frac{1}{2}\sqrt\frac{\tau s}{m^2}+\sqrt{\frac{\tau s}{4m^2}-1}\right],
\qquad t=\tau\sqrt{1-4\frac{m^2}{s\tau}}
\ee
Expressions (\ref{skkf},\ref{porcu},\ref{porcu2}) can be fetched into (\ref{lumia}) in order to obtain the cross section value for the $\foure$ process.
The price we pay is that it is no longer possible to obtain 
an analytic expression, and the cross section is evaluated numerically: we call this approximation ``numerical EPA''.  Contrarily to the leading log EPA, the numerical cross section is a much better approximation for the tree level cross section. This can be seen in Fig.\ref{fig:ImprovedEPA}, where we can see that the numerical EPA approximation (blue dots) is very close to both the CalcHEP numerical value (red dots) and to the analytic approximation of eq. (\ref{llepa4l}). The numerical EPA approximation will be very useful for the $6\ell$ case (see Par. \ref{par:6l}), where no numerical calculation can be performed and no analytic approximation was available before the writing of this paper.

\section{$e^+e^- \to e^+e^-e^+e^-e^+e^-$ six leptons in the Equivalent Photon Approximation and beyond\label{par:6l}}
\label{Sec:6lept}

The cross section of the SM process $\sixe$ is very hard to calculate due to the high number of particle in the final 
states producing several thousand of Feynman diagrams which contributes to the amplitude. Our attempt to obtain a numerical result for the cross section
failed using both CalcHEP\cite{Belyaev:2012qa} and MadGraph\cite{Alwall:2014hca} tools.  A formula to estimate the order of magnitude of the cross section is proposed in Eq. 12 of \cite{Batell:2009yf}:

\begin{equation}
\sigma_{e^+e^-\to 3(e^+e^-)}  \approx \frac{\alpha^6} {\pi^3 m_e^2} \left( \log\frac{s}{m_e^2} \right)^4
\label{eqn:6l_Pospelov}
\end{equation}

A slightly more precise estimate can be obtained by using the Leading Log EPA approximation by computing the cross section $\sigma(\gamma \gamma \to e^+e^-e^+e^-)$
as proposed by \cite{Budnev:1974de}:

\begin{equation}
\sigma_{e^+e^-\to 3(e^+e^-)}  \approx \frac{\alpha^2} {6\pi^2} \sigma_{(\gamma \gamma \to e^+e^-e^+e^-)} \left( \log\frac{s}{m_e^2} \right)^4
\label{eqn:6l_Cheng}
\end{equation}
where $ \sigma_{(\gamma \gamma \to e^+e^-e^+e^-)}$ can be estimated using \cite{Cheng:1970ef}.
This expression coincides with (\ref{llEPA6l}), that we derived using Leading Log EPA. Like in the $4\ell$ case, we expect  Eq. (\ref{eqn:6l_Cheng}) to be a very poor approximation of the tree level cross section. Unlike the 
4$\ell$ case, in the present literature there is no full analytic expression nor numerical calculation for the 6$\ell$ case. However we expect the Numerical EPA approximation to be a much better estimate of the tree level cross section than Eq. (\ref{eqn:6l_Cheng}). 

Using full numerical integration we derived a new expression for the cross section for the process $\sixe$ down to the constant term: 
\begin{equation}
\sigma_{EPAfull}=\frac{\alpha^2} {6\pi^2}\sigma_{(\gamma \gamma \to e^+e^-e^+e^-)} \left(\log ^4\left(\frac{s}{m^2}\right)+\textrm{A} \log ^3\left(\frac{s}{m^2}\right)+\textrm{B} \log ^2\left(\frac{s}{m^2}\right)+\textrm{C} \log \left(\frac{s}{m^2}\right)+\textrm{D}\right)
\label{eqn:6l_FullEPA}
\end{equation}
where the constant coefficients have the following values:
\begin{equation}
A \sim -11.9 \qquad
B \sim  22.62 \qquad
C \sim  143.5 \qquad
D \sim  -521.1
\end{equation} 
\begin{figure}
\begin{center}
\includegraphics[width=9 cm]{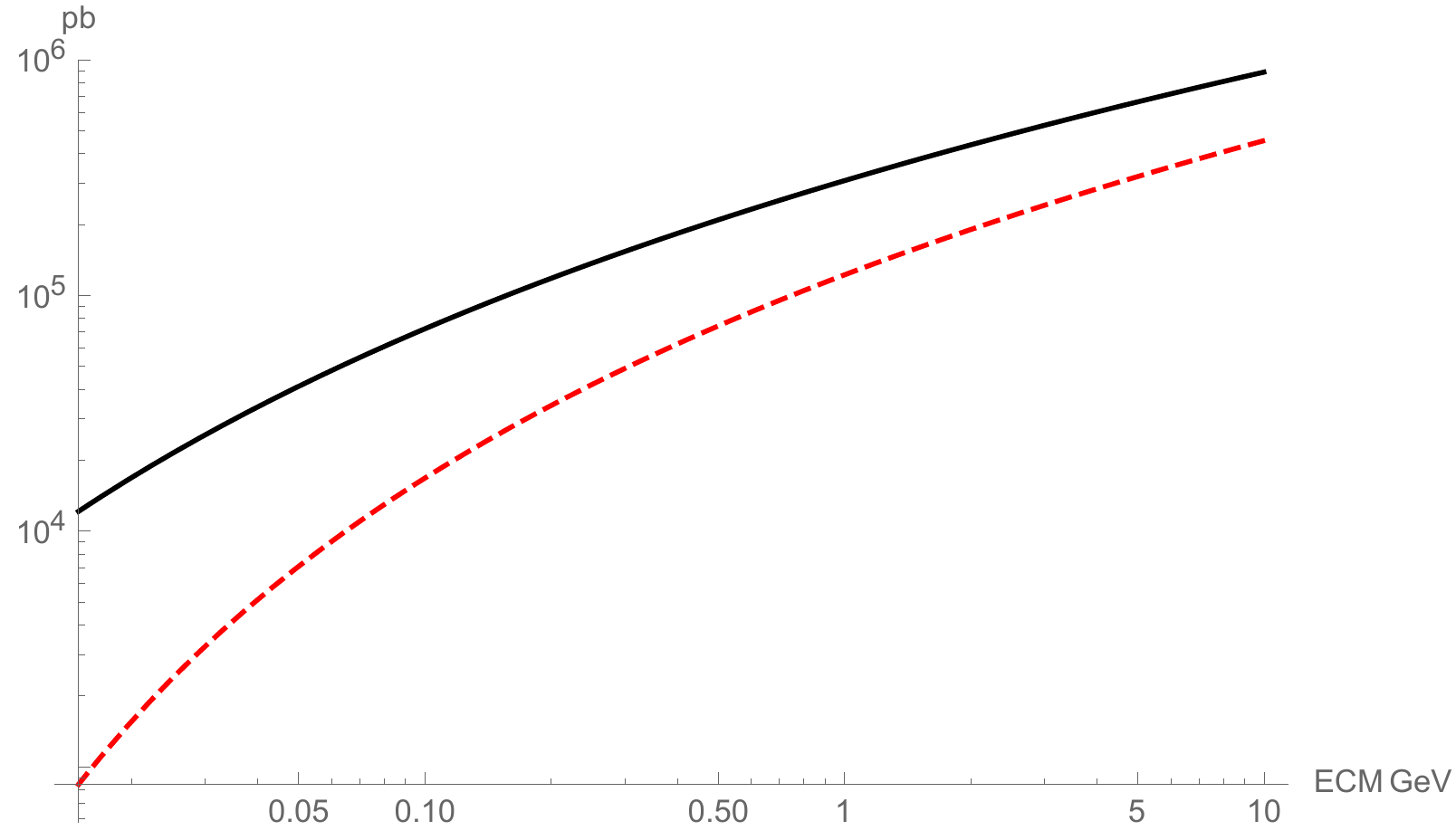}\\
\caption{{\em Leading Log EPA (in black) and Numerical EPA (red dashed) approximations for the tree level  $e^+e^-\to e^+e^-e^+e^-e^+e^-$ cross section.}}\label{fig:eeto6l}
\end{center}
\end{figure}

Like in the case of Eq. (\ref{eqn:4lBGMS_full}), accidental cancellation due to high values of the low power logarithmic term's constants is observed. In this case negative high
values of the constants $A$ and $D$, compared to the log scaling factor at low energies $<10$, strongly reduce the value of the total cross section, compared to the leading log term estimate.
In Fig. \ref{fig:eeto6l} the Leading Log EPA, Eq. (\ref{eqn:6l_Cheng}), and the numerical EPA cross sections, Eq. (\ref{eqn:6l_FullEPA}), are compared as function of the center of mass energy; like in the 4$\ell$ case, the leading log 
expression grossly overestimates the tree level cross section at low CM energies, while the disagreement decreases at high energy due to increased value of the Log term.
A key ingredient in the absolute value of the cross section common to the two parameterisation is the $\sigma_{(\gamma \gamma \to  4\ell)}$, assumed to be constant as function of the CM energy \cite{Budnev:1974de}.


\section{Phenomenology of Dark Higgs at low energy}

The authors of \cite{Batell:2009yf}  consider a minimal extension of the SM
by adding to the SM a $U(1)_D$ gauge boson $A'$ and a single complex scalar 
Higgs' field $\phi$ responsible for spontaneous symmetry breaking. 

They assume that any additional particles, in particular possible dark
matter candidates, are heavy compared to the $A'$ and $h'$ mass scales. 
All interactions with the SM and the secluded sector proceed
through kinetic mixing of $U(1)_D$ with the Standard Model photon.
We can neglect mixing with the Z boson and Standard Model Higgs boson which 
will be irrelevant at the energy scale under scrutiny for this paper.

The Lagrangian containing the physical Higgs' field $h'$ takes
the form:
\begin{equation}
\mathcal{L}=-\frac{1}{4}A^2_{\mu \nu}+\frac{1}{2}m^2_{A'}+\frac{1}{2}(\partial_{\mu} h')^2++\frac{1}{2}m^2_{h'}h'^2+\mathcal{L}_{int}
\end{equation}

With respect to the rich phenomenology described in \cite{Batell:2009yf} at low 
energy the situation is much simplified. Neglecting the possible decay to 
neutrinos, which is extremely suppressed at this energy, the only possible $A'$ 
decay is in $e^+e^-$ pairs. The decay width has the form:
\begin{equation}
    \Gamma_{A'\to ee}=\frac{1}{3}\alpha\epsilon^2m_{A'}\sqrt{1-\frac{4m^2_e}{m^2_{A'}}}\left(1+\frac{2m^2_e}{m^2_{A'}} \right)
\end{equation}
and for low energy and $\epsilon>1\times10^{-4}$ the decay $A\to e^+e^-$ is 
prompt.

Concerning the $h'$ decays only few possibilities exist. If the mass hierarchy is 
such that $m_{h'}>2m_{A'}$, the dominant decay is to a pair of on shell $A'$ 
($h'\to A'A'\to 2(e^+e^-)$).
In this scenario the decay is always prompt because the lifetime only depends on 
$\alpha_D$ which is supposed to be large compared to $\epsilon$.   
For masses $m_{h'}<2m_{A'}$ decays to off-shell $A'$ are also possible but, due to
the suppression of the 4 body decay $h'\to A'*A'*\to 2(e^+e^-)$, the dominant 
decay proceeds through triangle graphs to an $e^+e^-$ pair\cite{Batell:2009yf}. 
\begin{equation}
\tau(h'\to e^+e^-) \sim \frac{\alpha_D \alpha^2 \epsilon^4 m_{h'}}{2\pi^2} \frac{4m_e^2}{m_{A'}^2}   
\end{equation}
The lifetime in this case is extremely long and the Dark Higgs will be stable on the experiment scale 
producing missing energy in the final state.
Under the muon production threshold the possible dark sector final states induced
by $h'$ produced in association with an $A'$ are either $3(e^+e^-)$ or $(e^+e^-)$
+ missing energy.


\subsection{Dark Higgs Production cross sections}

One of the few $h'$ production processes is the so-called Higgs-strahlung, $e^{+}e^{-} \to h'A'$, which has an amplitude that is suppressed by just a single power of the kinetic mixing angle and can therefore readily occur for $\epsilon \sim O(10^{-2} - 10^{-3})$.
This production mechanism is similar to traditional Higgs-strahlung in the SM but in this case the $h'$ is produced in association with a $A'$ instead of a SM photon.
The total cross section for the Higgs-strahlung process reads \cite{Batell:2009yf}:

\begin{equation}\label{Eqn:DHCross}
\sigma_{e^{+}e^{-} \rightarrow A'h'} = \frac{\pi \alpha \alpha_D\epsilon^{2}}{3s} \left( 1 - \frac{m^{2}_{A'}}{s} \right)^{-2} \sqrt{\lambda \left( 1, \frac{m_{h'}^{2}}{s}, \frac{m^{2}_{A'}}{s} \right) }
\times \left[ \lambda \left( 1, \frac{m_{h'}^{2}}{s}, \frac{m^{2}_{A'}}{s} \right) + \frac{12m^{2}_{A'}}{s} \right] 
\end{equation}
where $\alpha_D$ is the coupling  between $h'$ and $A'$ and where $\lambda(a,b,c)\equiv a^2+b^2+c^2-2ab-2ac-2bc$.  

For reasonable values of the kinetic mixing parameter $\epsilon$, the cross section is quite large compared to the SM 6e. In fact the Higgs-strahlung cross section only pays an $\epsilon^2\alpha\alpha_D$ suppression compared to the $
\alpha^6$ of the concurrent SM process. For fixed values of the $m_{h'}$ and $m_{A'}$ the production cross section scales as 1/s which disfavour the production ~10 MeV scale particles at high energy colliders.  

\begin{figure}[htb]
\begin{center}
\includegraphics[width=8 cm]{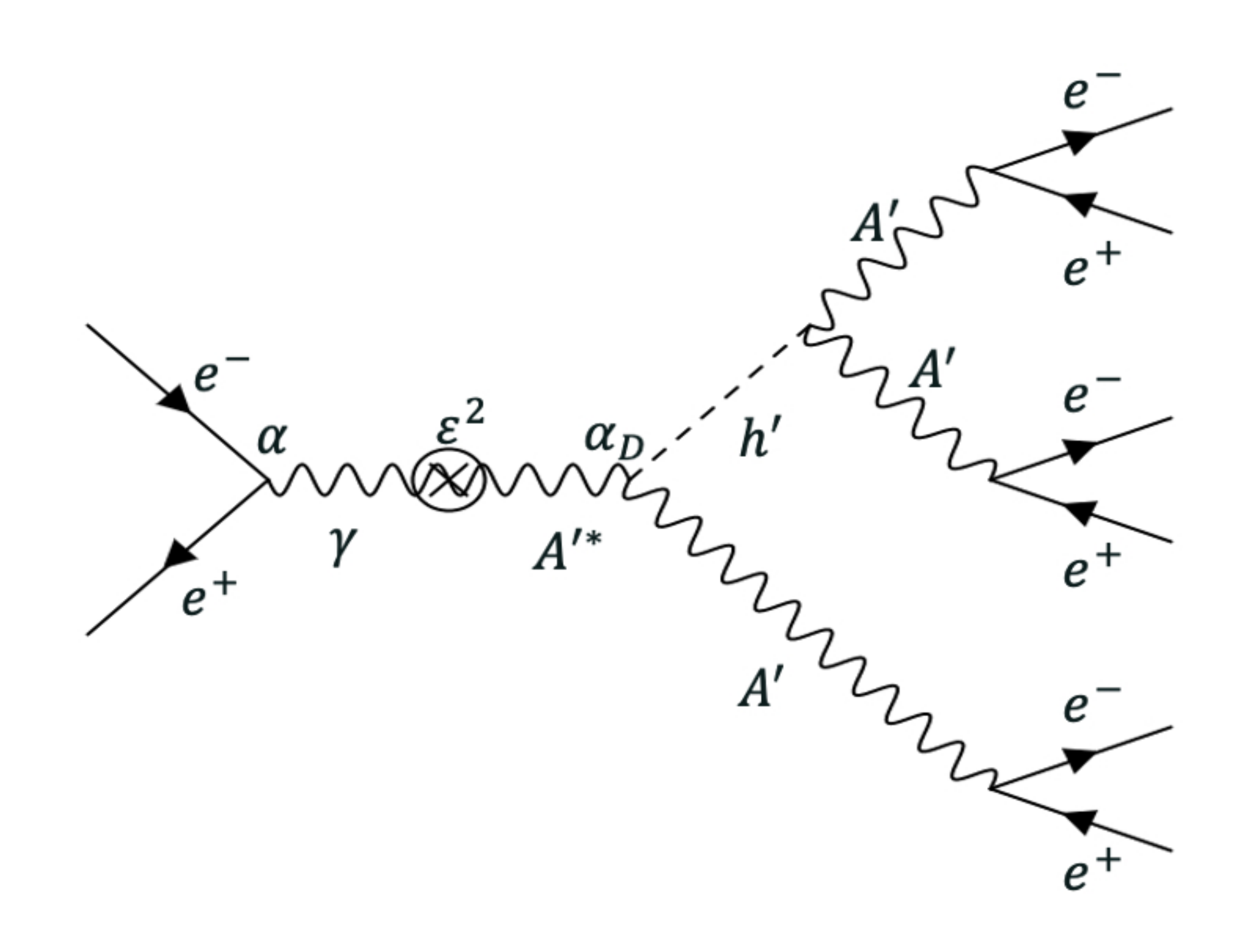}\\
\caption{{\em Feynman diagram for the Higgs-strahlung process producing a six leptons final state.}}
\label{fig:hstrahalung}
\end{center}
\end{figure}





\subsection{SM 6-leptons and Dark Higgs decay comparison}

We pointed out that $\sixe$ final state in $e^+e^-$ collisions can be produced by two competing mechanisms:
the beyond the SM process $e^{+}e^{-} \to h'A' \to 3(e^+e^-)$, shown in Fig. \ref{fig:hstrahalung} and the Standard Model process 
$\sixe$. In this section we will estimate the relative contribution to experimental total cross section $e^{+}e^{-} \to 3(e^+e^-)$ which is crucial
in searching for the BSM signal mediated by $h'$. We will only consider the scenario with all particles on shell given by the mass hierarchy:
$m_{h'}>2m_{A'}$, $m_{A'}>2m_e$. In this conditions the kinematic is maximally
different with respect to $\sixe$ SM process, and the $h'$ decays always prompt.

The physics of the two processes suggest that, in particular at low energy, the BSM process (signal) will produce a non negligible contribution to the total cross section.
In fact the SM $\sixe$ cross section (background) grows as $\sim \left(\log \frac{s}{m_e^2}\right)^4$ while the Higgs-strahlung cross sections drops as $1/s$, resulting in a signal 
over background scaling :
\begin{equation}
\frac{S}{B} \approx \frac{m_e^2}{s \left(\log\frac{s}{m_e^2} \right)^4}.
\label{Eqn:SBdrop}
\end{equation}
As we already pointed out in section \ref{Sec:6lept}, due to accidental cancellation the LLEPA approximation grossly overestimates the $\sixe$ SM cross section, which is not dominated by the $\log^4$ term,  
and therefore the S/B scaling will actually be smaller than naively expected using Eq. \ref{Eqn:SBdrop}.
\begin{figure}[htb]
\begin{center}
\includegraphics[width=8 cm]{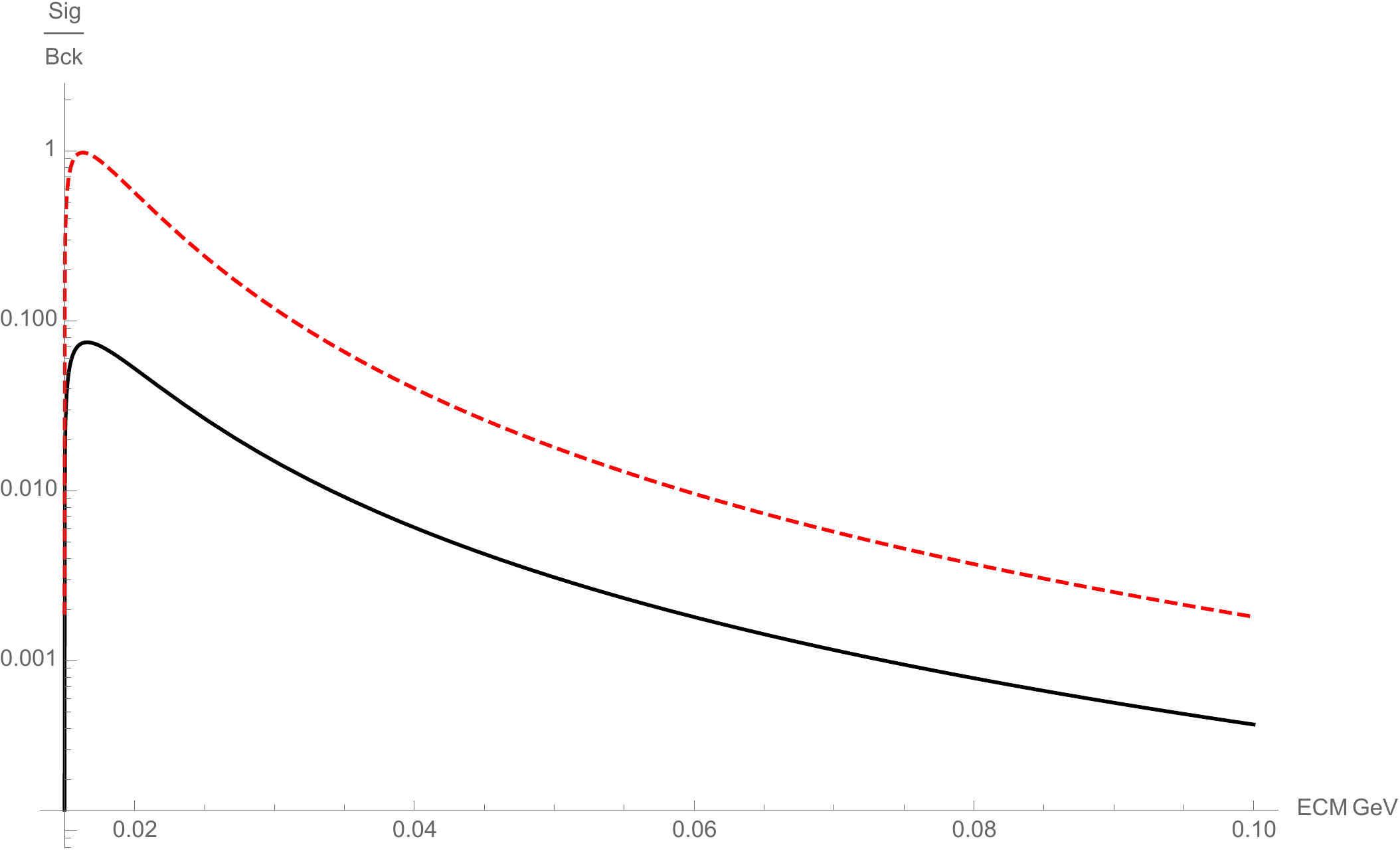}\\
\caption{{\em Black line: S/B ratio using Leading Log EPA approximation. Red dashed line: S/B ratio using numerical EPA approximation.}}
\label{fig:SBratio}
\end{center}
\end{figure}

Nevertheless we also observed that the actual value of the $\sixe$ cross section is much smaller than expected, providing a much higher starting value of the S/B ratio.
The ratio $\sigma_{h'A'}/\sigma_{6e}$ is shown in Fig. \ref{fig:SBratio} for center of mass energies ranging from 0.02 GeV to 0.1 GeV setting $m_{h'}$=10 MeV, $m_{A'}$=5 MeV, $\epsilon^2=1\times10^{-6}$ and $\alpha_D=0.1$ in Eq.
\ref{Eqn:DHCross}. 

The value of $\epsilon^2$ has been chosen small enough to evade experimental constraints coming from visibile dark photon searches\cite{Battaglieri:2017aum}.
It's interesting to point out that, after the new precise calculation of $\sigma_{\sixe}$ proposed in this work, the S/B ratio (red line) is increased by roughly one order of magnitude 
compared to previous estimates (black line). Up to a CM energy $\sim$30 MeV the contribution from $h'$ to the total cross section exceeds 10$\%$, 
and will therefore be observable even by just measuring the total $\sixe$ cross section. When all the particle are on mass shell the value of the $\lambda$ factors in Eq. (\ref{Eqn:DHCross}) varies
just by small factors changing the values of $m_{h'}$ and $m_{A'}$ having it's maximum for $m_h'\sim m_{A'}$.

\subsection{Background suppression}
In this section we would like to point out that appropriate experimental cuts on the final $\sixe$ state can strongly enhance the S/B ratio. Let us begin by noticing that the distribution of the virtual photons emitted by the initial positron and 
electron is peaked in the region of small energy and small angle. This can be seen from Eq.\ref{structure} which behaves like $\sim\frac{d\kp^2}{\kp^2}\frac{dx}{x}$, $x$ being the fraction of the photon's energy and $\kp$ the photon's 
transverse momentum with respect to the parent electron (or positron). 
This implies that the initial electron (positron) emits a final electron (positron) which is collinear and has about the same energy of the initial electron (positron). 

In other  words, most of the final events are characterised, in the c.m. frame, by a pair electron-positron emitted back to back  and with an invariant mass close to the initial c.m. energy $\sqrt{s}$. This signature is very different from that 
of the signal event:  in this case two mass shell bosons ($h'$ and $A'$) are created and, broadly speaking, the energy is evenly distributed between the final $e^+ e^-$ pairs.  Moreover, three $e^+e^-$ pairs invariant masses  reconstruct 
the $A'$ mass. 

In the case of PADME things happen in the lab frame, and in this case we will have in most of the events a final positron collinear to the beam and carrying almost all of the available energy. Also in this case then, we 
expect the signal signature to be very different from the background one.
To precisely compute the kinematical distributions of the events in PADME goes beyond the scope of the present paper, but it should be clear that such a study can only result in a S/B ratio better than the one shown in Fig. \ref{fig:SBratio}.

\section{Multi lepton perspectives in PADME}

In this section we try to derive the possible perspective for observing multi lepton final states and eventually constrain the $h'$ parameter space with the PADME experiment at the INFN Laboratori Nazionali di Frascati \cite{Raggi:2014zpa}\cite{Raggi:2015gza}. In present literature no observation are reported for $\sixe$ or $\foure$ processes and no cross section is available which makes their observations appealing. 

PADME is a positron on target experiment with a $\sim$ 500 MeV energy positron beam impinging on a diamond target. Due to the low beam energy PADME has the opportunity to search for dark higgs 
in the mass region from few to 15 MeV, very difficult to achieve for high energy colliders due to the overwhelming associated $\sixe$ background. The experiment is currently taking data for Run II at the DA$\Phi$NE linac and aims to 
collect $\sim 1\times10^{13}$ positron on target collisions. 
PADME is a fixed target experiment with a diamond target of d=100$\mu$m.
The experiment luminosity can be computed using the formula:
\be
L_{inst}=
\frac{I_b}{e}N_A\frac{Z \rho d}{A}
\ee
The relevant quantities are Z= 6, density of diamond, $\rho$=3.51 g/$cm^3$, and A = 12.01 g carbon's gram-molecular weight.
The $I_b/e$ during PADME Run II, corresponds to 49 bunches/s of $\sim27000$ positrons leading to $1.3\times10^6$ positrons on target per second. 
The resulting instantaneous luminosity of the experiment is $\sim1.4\times10^{-8}$ pb/s. Considering one year of data taking corresponding to $1\times10^7$s of effective
experiment's running time, the integrated luminosity will be $\sim$ 0.14pb $^{-1}$/Y.
According to the ratio of Fig. \ref{fig:SBratio}, at the CM energy PADME is operating ($\sqrt{2m_eE_{Beam}}\sim22 MeV$), the existence of an $h'$ will produce very strong 
enhancement of the SM $\sixe$ cross section even with $\epsilon^2$ values as small as $10^{-6}$. 
To estimate production rates of $\sixe$  and $h'$ at PADME, we can use Eq. \ref{Eqn:DHCross} and Eq. \ref{eqn:6l_FullEPA} to compute the actual values of the signal and background cross sections at $E_{Beam}$=500 MeV.
\begin{itemize}
\item The dark higg's production cross section $\sigma_{h'A'} \sim 1000$ pb\\
	One year of PADME data taking with the instantaneous luminosity of Run II will allow to produce $\sim$150 $h'$ mediated $e^+e^-$ in six leptons final states.  
\item $\sigma_{\sixe} \sim1500$ pb \\
          One year of PADME data taking with the instantaneous luminosity of Run II will allow to produce $\sim$200 SM $\sixe$ events.  
\end{itemize}

Even a modest BG rejection obtained with the present PADME veto system, will allow the dark sector to become the dominant process.
A crucial aspect to understand the experimental acceptance, due to minimum track momentum in the PADME
spectrometer ($\sim$ 50 MeV), will be to derive the momentum distributions of tracks coming from the two concurring processes,
which we are at present not able to derive for $\sixe$. Nevertheless we expect that a small fraction of $\sixe$ event will have a minimum track
energy higher than 50 MeV, reducing their acceptance with respect to $h'$ mediated $6\ell$ final states.

Important information about the dynamic of the $\sixe$ process could be derived from the observation of $\foure$. 
A first evidence of this process was obtained in the early `70 at Laboratori Nazionali di Frascati\cite{Bacci:1972tk} without being able to obtain a cross section measurement. 
In this case the cross section is much higher ($\sim \alpha^2 =10000$), as it is the expected to be the experimental acceptance due to higher average track momentum.
Computing the cross section value with Eq. (\ref{eqn:4lBGMS_full}) at PADME energies we get 1.2$\times10^8$ pb leading to order 
$10^7$ $\foure$/y at PADME. Even with a small experimental acceptance PADME has the possibility of observing for the 
first time this process at energies below 1GeV already in Run II data.  




\section{Conclusions}

In the present paper we revised current EPA based calculations of the SM cross section for the processes $\foure$ and $\sixe$, in order to compare them to possible dark sector 
contributions. We pointed out that, due to accidental cancellations, the leading logarithm approximation for the cross section it is a gross estimate at energies below 10 GeV,
relevant for both fixed target experiments and collider one like Belle II.
We confirmed, using complete tree level calculation and improved analytic EPA based calculations, the results for the $\foure$ obtained by authors of \cite{Budnev:1974de} with a full EPA based calculation.
Using our new approach we have been able to derive for the first time an analytic cross section of $\sixe$ including all the logarithmic terms. 
Comparing the result with the Leading Log EPA approximation we discovered that the cross section is up to 1 order of magnitude lower with respect to previous estimates. 
Concerning dark sector contributions, as consequence, we observed that the contribution of $h'$ mediated $6\ell$ final state its dominant at low energies, even for small values of the $\epsilon$ mixing parameter.
The potential in measuring for the first time the $\foure$ and $\sixe$ cross section and constraining the dark sector observable with the PADME experiment is also explored. 


\appendix
\section{The Equivalent Photon Approximation (EPA)\label{appA}}

In this paper we consider QED processes of the kind $e^+e^-\to e^+e^- X$, where $X=e^+e^-$ or $X=e^+e^-e^+e^-$. These processes are hard to calculate even at tree level, especially the second  one that features a huge number of 
Feynman diagrams and a six-body phase space; in fact no tree level analytical or numerical  calculation is available for this process. It is however possible to use the Equivalent Photon Approximation, whose origin dates back to a long 
time ago \cite{fermi,weisz},  that has been used also in the context of weak interactions \cite{Dawson}, and that in the case of strong interactions is better known as Altarelli-Parisi equations for the evolution of structure functions \cite{Altarelli:1977zs}. In this Appendix we describe this approximation and estimate the uncertainty with which the cross sections of our interest can be calculated. 
\begin{figure}
\begin{center}
\includegraphics[width=8 cm]{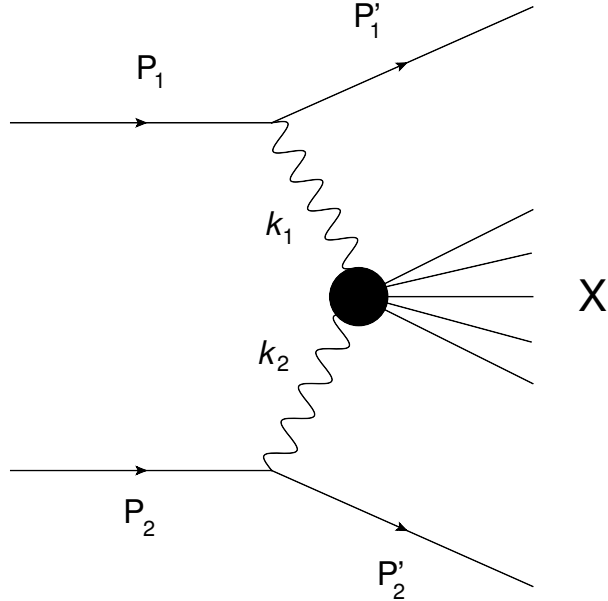}\\
\caption{{\em Diagram of the Equivalent Photon Approximation for the process
$e^+ e^-\to e^+e^-X$, where  $X $ is a generic multiparticle state. An electron ($P_1$) and a positron ($P_2$) emit one photon each; the photons scatter through the process $\gamma(k_1) \gamma(k_2)\to X$. In this paper we consider $X= e^+e^-$ and $X= e^+e^-e^+e^-$.}}\label{fig:EPA}
\end{center}
\end{figure}

In order to understand EPA, it is crucial to observe that the photon propagators become singular when the emitted photon is collinear with emitting electron. Indeed, emitting indices for the sake of clarity and defining  $k=(\omega,\kvet)$, $P=(E,\pvet)$,
$P'=(E',\pvet')$, $\pvet\cdot\kvet=|\pvet||\kvet|\cos\theta$ we have\footnote{Note that $-k^2$ is positive defined, i.e. $k^2<0$}:
\begin{equation}\label{kk}
k^2=2m^2-2EE'+2EE'\sqrt{1-\frac{m^2}{E^2}}\sqrt{1-\frac{m^2}{E'^2}}\cos\theta
\end{equation}
where $m$ is the electron mass. Since  we are interested in  the limit
$E,E'\gg m$, we can Taylor expand  (\ref{kk}) in the small parameter $m^2$, obtaining:
\be
-k^2=-2EE'(1-\cos\theta)+(\frac{E}{E'}\cos\theta+\frac{E'}{E}\cos\theta-2) m^2+{\cal O}(m^4)
\ee
As $\theta\to 0$ the first term in the expansion drops down and $-k^2$ is of the order of the small parameter $m^2$. If also $m$ goes to 0, then the singularity is exposed since  $\frac{1}{k^2}\to\infty$. In other terms, $m$ cuts off the $\theta=0$ singularity; for this reason this is called a `mass singularity'. In any case, the presence of this singularity enhances the contribution of the EPA diagram of fig. (\ref{fig:EPA}) with respect to the other diagrams contributing to the process; from the discussion above it is clear that the enhancement grows bigger and bigger as the c.m energy of the process grows, so that the EPA approximation becomes a better one  at high energies. In the following we shall see that the enhancement is logarithmic, i.e. proportional to $L=\log\frac{E}{m}$.

In the region that gives the largest contribution, which is the collinear region, where the emitted photon is (almost) collinear to the emitting electron, we can write $k=xP+\kp$. Here $x$ is the fraction of the electron momentum carried by the gluon, while $\kp$ is the component of $\kvet$ perpendicular to $\pvet$. For the reasons stated above, $\kp$ is `small', $|\kp|\ll E$. With these notations,  the Equivalent Photon Approximation then amounts to writing (see fig. \ref{fig:EPA}):
\be\label{mainEPA}
\sigma({e^+e^-\to e^+e^- X};s)=\int dx_1 dx_2 f_{\gamma e}(x_1)
 f_{\gamma e}(x_2)\sigma({\gamma\gamma\to X};Q^2=x_1x_2s)
\ee
where $s$ is the invariant mass of the initial state, while $Q^2$ is the invariant mass of the $\gamma \gamma$ pair. There are a few simplifying hypotheses underlying the EPA, namely:
\begin{itemize}
\item
Eq. (\ref{mainEPA}) is a probabilistic model in which the emission of photons with distribution $f_{\gamma e}(x_i)$ is factorized with respect to the photons scattering. However, in order to do the computation, one should consider all the relevant Feynman diagrams, add them them including interferences. The rationale behind the simplified expression is that the contribution in fig. \ref{fig:EPA} is enhanced by the two photon propagators, as discussed above.
\item 
The $\sigma({\gamma\gamma\to X};Q^2=x_1x_2s)$ cross section is calculated with onshell photons ($k_i^2=0$) while the photons are virtual in the diagram. moreover the invariant mass of the subprocess $\gamma\gamma\to X$ is calculated as $Q^2=x_1x_2$, thus neglecting the transverse components of the photons' momenta.
\end{itemize}
While it is hard in general to assess the precision with which the EPA approximates the `true' value for the cross section, at the end of this section we estimate such a precision for the processes of relevance in this paper. 

The structure function $f_{\gamma e}$ can be calculated in QED to give \cite{boh}:
\be
\label{structure}
f_{\gamma e}(x)=\frac{\alpha}{2\pi}\int_{(\kp^2)_{min}}^{(\kp^2)_{max}}\frac{d\kp^2}{\kp^2}P_{\gamma e}(x); \qquad P_{\gamma e}(x)=\frac{1+(1-x)^2}{x}
\ee
Notice that one can apply the approximations leading to ({main:EPA},\ref{structure}) to very different physical situations. For instance, provided the energy is sufficiently high, the electron can emit a weak gauge boson instead of a photon \cite{Dawson}. In this case one talks about Equivalent Boson Approximations and the changes with respect to the QED case considered here are: the substitution of $\alpha$ 
with the weak coupling constant $\alpha_W$ and the appearance of a structure function for longitudinal bosons, absent in the case of photons\footnote{The  ElectroWeak case for more general cases has been studied in \cite{Ciafa}}. The initial particle can be for instance a proton instead of an electron, and one then has distributions for quarks and gluons. The $P_{gq}$ Altarelli Parisi splitting function where $g$=gluon and $q$=quark is, in fact, identical to the $P_{\gamma e}(x)$ appearing in (\ref{structure}). There are, however, a couple of notable differences between QED and QCD. In the case of QED, the extrema $(\kp^2)_{min},{(\kp^2)_{max}}$ are determined by kinematics and this allows for a calculation of $f_{\gamma e}$, as we show below. In the case of QCD however, while
${(\kp^2)_{max}}$ is of the order of the invariant mass $Q^2$ of the subprocess,  $(\kp^2)_{min}$ is given by an infrared cutoff $\mu^2$ of the order of the scale $\Lambda_{QCD}$ below which the perturbative calculation is senseless due to uncalculabel nonperturbative effects. The integration over $\kp^2$ gives  is then  $\log\frac{Q^2}{\mu^2}$ and depends on the arbitrary scale $\mu$ so that $f_{gq}$ is not perturbatevely calculable in QCD. Nevertheless, what (\ref{structure}) provides is an information on the dependence of the structure function on the $Q^2$ of the process. Then, one can measure a given structure functions at a `low' scale
$Q_0^2$ and use this equation to determine the structure function at a `high' scale $Q^2$. This task is performed by the Altarelli-Parisi equations, that are an all-order extension of (\ref{structure}) in the form of integro-differential equations that take into account large logaritms
$\log\frac{Q^2}{Q_0^2}$ to all orders in perturbation theory. And here is another difference with the EPA approach, which considers only terms of the first order in $\alpha$.

Getting back to the EPA, since the `partonic' cross section $\sigma_p(\gamma\gamma\to X)$  depends on the combination $x_1x_2$, it is convenient to change      variables $x_1,x_2\to x=x_1,\tau=x_1x_2$ in (\ref{mainEPA}), obtaining;
\be\label{lumi}
\sigma(s)=\int_\epsilon^1d\tau{\cal L}(\tau)\sigma_p(\tau s)\qquad
{\cal L}(\tau)=\int_\tau^1\frac{dx}{x}f_{\gamma e}(x)f_{\gamma e}(\frac{\tau}{x})
\ee
where $\epsilon$ is defined in terms of the threshold invariant mass $M_{th}^2$ of the partonic subprocess; for instance in the case of $\gamma \gamma\to e^+e^-$ we have $\epsilon\equiv \frac{M^2_{th}}{s}=\frac{4 m^2}{s}$. In the case of the processes considered in this paper, a careful examination of kinematical bounds entails:
 \be
 f_{\gamma e}(x)=\frac{\alpha}{2\pi}\log\frac{1}{x^2}\Rightarrow 
{\cal L}(\tau)\approx\frac{2\alpha^2}{3\pi^2}\frac{(\log\frac{1}{\tau})^3}{\tau}
 \ee
 where the value of the luminosity ${\cal L}$ has been obtained by fetching $f_{\gamma e}(x)$ into
 (\ref{lumi}) and neglecting terms of order $(\log\frac{1}{\tau})^2$ (leading log approximation). Using this luminosity and the known expression for $\sigma(\gamma\gamma)\to e^+e^-$ \cite{Budnev:1974de} we obtain:
 \be
 \sigma(e^+e^-\to e^+e^-e^+e^-)\approx \frac{28\alpha^4}{27\pi m^2}(\log\frac{s}{m^2})^3
 \ee
 so that the leading behaviour of the cross section is a growth like $L^3=(\log\frac{s}{m^2})^3$. The situation is different for $e^+e^-\to 6 leptons$: the luminosity is the same, but the `partonic' cross section $\gamma\gamma\to $4 leptons is approximately constant, differently from the case of $\gamma\gamma\to $4 leptons where the cross section decreases like $1/Q^2$ for high $Q^2$. We obtain:
 \be\label{llEPA6l}
\sigma_{2e\to 6e}(s)\approx\frac{\alpha^2}{6\pi^2}l^4\sigma_{\gamma\gamma\to 4e}=\label{giusta}
\frac{\alpha^6l^4}{3\pi^3 m^2}\frac{175 \,\zeta(3)-38}{72}
\approx 0.79\frac{\alpha^6l^4}{\pi^3 m^2}
\ee
 where the value for $\sigma_{\gamma\gamma\to 4e}$ has been taken from \cite{Cheng:1970ef}.


\begin{thebibliography}{100}
\bibitem{Batell:2009yf}
B.~Batell, M.~Pospelov and A.~Ritz,
Phys. Rev. D \textbf{79}, 115008 (2009)
doi:10.1103/PhysRevD.79.115008

\bibitem{Essig:2009nc}
R.~Essig, P.~Schuster and N.~Toro,
Phys. Rev. D \textbf{80}, 015003 (2009)

\bibitem{babar} BaBar Collaboration (Lees J. P. et al.), Phys. Rev. Lett., \textbf{108} (2012) 211801.

\bibitem{belle} Belle Collaboration (Jaegle I.), Phys. Rev. Lett., \textbf{114} (2015) 211801.



\bibitem{Budnev:1974de}
V.~M.~Budnev, I.~F.~Ginzburg, G.~V.~Meledin and V.~G.~Serbo,
Phys. Rept. \textbf{15}, 181-281 (1975)

\bibitem{Accardi:2020swt}
A.~Accardi, A.~Afanasev, I.~Albayrak, S.~F.~Ali, M.~Amaryan, J.~R.~M.~Annand, J.~Arrington, A.~Asaturyan, H.~Avakian and T.~Averett, \textit{et al.}
[arXiv:2007.15081 [nucl-ex]].



\bibitem{Aubert:2009af}
B.~Aubert \textit{et al.} [BaBar],
[arXiv:0908.2821 [hep-ex]].

\bibitem{Belyaev:2012qa}
A.~Belyaev, N.~D.~Christensen and A.~Pukhov,
Comput. Phys. Commun. \textbf{184}, 1729-1769 (2013)
doi:10.1016/j.cpc.2013.01.014
[arXiv:1207.6082 [hep-ph]].

\bibitem{Alwall:2014hca}
J.~Alwall, R.~Frederix, S.~Frixione, V.~Hirschi, F.~Maltoni, O.~Mattelaer, H.~S.~Shao, T.~Stelzer, P.~Torrielli and M.~Zaro,
JHEP \textbf{07}, 079 (2014)
doi:10.1007/JHEP07(2014)079

\bibitem{Cheng:1970ef}
H.~Cheng and T.~T.~Wu,
Phys. Rev. D \textbf{1} (1970), 3414-3415
doi:10.1103/PhysRevD.1.3414

\bibitem{Battaglieri:2017aum}
M.~Battaglieri, A.~Belloni, A.~Chou, P.~Cushman, B.~Echenard, R.~Essig, J.~Estrada, J.~L.~Feng, B.~Flaugher and P.~J.~Fox, \textit{et al.}
[arXiv:1707.04591 [hep-ph]].

\bibitem{Raggi:2014zpa}
M.~Raggi and V.~Kozhuharov,
Adv. High Energy Phys. \textbf{2014}, 959802 (2014)

\bibitem{Raggi:2015gza}
M.~Raggi, V.~Kozhuharov and P.~Valente,
EPJ Web Conf. \textbf{96}, 01025 (2015)

\bibitem{Bacci:1972tk}
C.~Bacci, et al.
Lett. Nuovo Cim. \textbf{3S2}, 709-714 (1972)
doi:10.1007/BF02824344


\bibitem{boh}
Y.~L.~Dokshitzer,
Sov. Phys. JETP \textbf{46} (1977), 641-653;
V.~N.~Gribov and L.~N.~Lipatov,
Sov. J. Nucl. Phys. \textbf{15} (1972), 438-450
IPTI-381-71.
\bibitem{fermi}
E. Fermi, Z. Phisik 29 (1924) 315.
\bibitem{weisz}
C.~F.~von Weizsacker,
Z. Phys. \textbf{88} (1934), 612-625
doi:10.1007/BF01333110;
E.~J.~Williams,
Kong. Dan. Vid. Sel. Mat. Fys. Med. \textbf{13N4} (1935) no.4, 1-50.

\bibitem{Dawson}
S.~Dawson,
Nucl. Phys. B \textbf{249} (1985), 42-60
doi:10.1016/0550-3213(85)90038-0
\bibitem{Altarelli:1977zs}
G.~Altarelli and G.~Parisi,
Nucl. Phys. B \textbf{126} (1977), 298-318
doi:10.1016/0550-3213(77)90384-4
\bibitem{Ciafa}
P.~Ciafaloni and D.~Comelli,
JHEP \textbf{11} (2005), 022
doi:10.1088/1126-6708/2005/11/022
[arXiv:hep-ph/0505047 [hep-ph]].







\end{thebibliography}
\end{document}